# Designing a Reliable Inland Waterway Transportation Network under Uncertainty

**Amin Aghalari, Farjana Nur, Mohammad Marufuzzaman**
**Department of Industrial & Systems Engineering Mississippi State University, Starkville, MS, 39759-9542**

**Stephen M. Puryear**
**Center for Advanced Vehicular Systems Extension (CAVSE)**
**Mississippi State University, 153 Mississippi Parkway Canton, MS, 39046**

## Abstract

Inland waterway transportation network significantly supports the overall freight transportation of the nation. In order to ensure efficient and timely commodity transportation through this network, this study aims at developing a reliable inland waterway transportation network considering the interactions between different transportation entities along with considering the uncertain commodity supply and unpredictable waterway conditions over time. A capacitated, multi-commodity, multi-period, stochastic, two-stage mixed-integer linear programming (MILP) model is proposed to capture this stochastic, time variant behavior of the water-depth along any link of the inland waterway under consideration. Additionally, we proposed a parallelized hybrid decomposition algorithm to solve the real-life test instances of this complex NP-hard problem. The proposed algorithm is capable of producing high quality solutions within a reasonable amount of time. Further, a case study is demonstrated for the Southeast region of the United States and a number of managerial insights are drawn that magnifies the impact of different key input parameters on the overall inland waterway transportation network under consideration.

## 1. Introduction

The inland waterway ports are one of the greatest contributors of the U.S. economy that tremendously supports the overall freight transportation of the nation while creating more than 250,000 job opportunities (both direct and indirect) nationwide. These ports contribute to the rural, industrial, and agricultural development of the nation and provide the most economical and environmentally friendly option compared to other transportation modes. However, the inland waterways are greatly suffering from many waterway and infrastructure related issues among which water level fluctuation, can be considered as the most crucial one. To exemplify, in spring 2011, a severe flood affected the inland waterway system of the U.S. that caused a total damages of approximately $8.5 billion. While in the very next year waterways experienced severe drought causing a number of barges to run aground. Considering the severity and frequency of this vital waterway issue, developing reliable model for inland waterway transportation system accounting waterway specific issues becomes very crucial to retain and support the inland waterway freight transportation of the country. Inland waterway ports possess some distinguished features over seaports. For example, these ports are located near smaller water bodies, only handle shallow draft vessels, transport mainly agricultural products, and many others. Due to the presence of these unique features, the available seaport literature cannot be directly applied to model the inland waterway port activities. Therefore, sophisticated optimization models need to be developed to best capture these characteristics and address all issues that disturbs this cost efficient, environmentally friendly transportation mode.

To date many researchers has studied diversified seaport-related problems; but, compared to seaports, inland waterway ports did not receive much attention from the research community. Some studies considered deep draft inland ports that are capable of handling deep draft vessels. The issues considered in those studies are barge and towboat routing and repositioning [2], berth layout design and allocation [3], port disruption [4], delays in locks and dams [5]. Other studies related to deep draft inland waterway ports include the consideration of port-specific economic analysis [6], optimal dredging scheduling and investment decisions [7], the efficiency of inland waterway container terminals [8], tug scheduling between seaport to inland ports [9], and few others. These studies detail the specifics of deep draft inland ports. However, Shallow draft inland ports that handle barges/vessels up to 9 feet only has not received much attentions from the research community. Very few studies consider inland waterway ports as a medium of transportation to design a supply chain network such as biomass supply chain (e.g., [10]), coal supply chain (e.g., [11]), grain supply chain (e.g., [12]), and many other application areas. However, almost no research has been conducted to date that captures the specific attributes of these set of ports while their contributions to the overall transportation system and economy are outstanding. Therefore, better understanding of shallow draft inland waterway ports is imperative to successfully design and manage a sound and efficient supply chain network. In order to address this literature gap, this study proposes a capacitated, multi-commodity, multi-period, two-stage stochastic mixed-integer linear programming model that jointly optimizes trip-wise towboat and barge assignment decisions along with different supply chain decisions (e.g., inventory management, transportation decisions) under uncertain waterway conditions in such a way that the overall supply chain cost is minimized. Realizing the computational complexity of this NP-hard problem, we develop a highly customized parallelized hybrid decomposition algorithm to solve large instances of our proposed optimization model in a reasonable amount of time. Additionally, we demonstrate a real-life case study that reveals important managerial insights about critical performance measures of the inland waterway transportation network.

## 2. Problem Description and Model Formulation

Figure 1 demonstrates a simplified inland waterway transportation network consisting of two origin ports and three destination ports. Consider a logistic network where set $I = \{1, 2, 3, ..., I\}$ be the set of origin ports and set $J = \{1, 2, 3, ..., J\}$ be the set of destination ports. A set of commodities $M = \{1, 2, 3, ..., M\}$ needs to be transported through the two tiers (origin ports and destination ports) of the network over a predetermined set of time periods $T = \{1, 2, 3, ..., T\}$. Subsets $I_j$ and $J_i$ are introduced, where, set $I_j$ represents the subset of origin ports connected to port $j \in J$ and $J_i$ represents the subset of destination ports connected to origin port $i \in I$. To account for different scenarios of commodity supply and water-level fluctuations, the scenario set $\omega \in \Omega$ is introduced where rw defines the probability of a given realization and $\sum_{\omega \in \Omega} \rho_\omega = 1$.

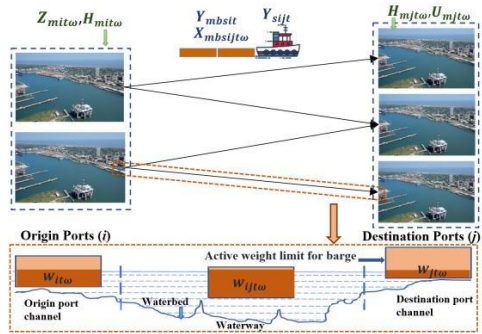

Figure 1: Illustration of a inland waterway transportation network

Origin port $i \in I$ is supplied with a stochastic amount of commodity $\phi_{mit\omega}$ of type $m \in M$ at time period $t \in T$ under scenario $\omega \in \Omega$. The procurement cost of commodity $m$ in port $i$ at time period $t$ is denoted by $\gamma_{mit}$. Based on demand these commodities are transported to the destination ports using a set of barges $B = \{1, 2, 3, ..., B\}$ and a set of towboats $S = \{1, 2, 3, ..., S\}$ through the inland waterway transportation network, where both towboats and barges can only be used one-sided and once in a period. Towboats are capacitated and we denote $\overline{\delta_s}$ and $\underline{\delta_s}$ to be the maximum and minimum number of barges that can be carried out by any particular towboat $s \in S$. The fixed loading and unloading commodity cost of carrying commodity $m \in M$ in barge $b \in B$ and the fixed cost of using a towboat $s \in S$ at any given time period $t \in T$ is denoted by $\eta_{mbt}$ and by $\psi_{st}$. Three additional parameters $\overline{w}_b$, $v_b$, and $\rho_m$ are introduced to represent the weight carrying capacity of barge $b$, volumetric capacity of barge $b$, and the density of commodity $m$, respectively.

The unit commodity ($m \in M$) transportation cost along arc $(i,j) \in (I,J)$ using any barge $b$ connected with towboat $s$ at time period $t$ is denoted by $c_{mbsijt}$. Parameters $a_{bit}$ and $a_{sit}$ represent the binary availability of barges and towboats at any port $i$ while their periodic maintenance is being considered. Each port $i \in I \cup J$ is equipped with inventories where maximum total of $\overline{h}_i$ commodity units can be stored with a holding cost of $h_{mit}$ for each commodity $m$ at time $t$. While storing commodity $m$ between two consecutive time periods, commodity deterioration can occur in a rate of $\alpha_m$. In order to define the active weight capacity of any barge between any specific port pairs at any given time parameters $w_{it\omega}$, $w_{jt\omega}$, and $w_{ijt\omega}$ are introduced. With $w_{it\omega}$ and $w_{jt\omega}$ we represent the maximum weight carrying capacity at port channel $i \in I \cup J$ and $w_{ijt\omega}$ denote the allowable weight that can be carried through the waterway between the respective ports $(i,j) \in (I,J)$ at time period $t$ under scenario $\omega$. Note that, due to sediment, silt, or mud accumulation in the waterbed the waterway depth different parts of waterway may vary in different time period of the year. With such accumulation being too high at any waterway segment the height of the waterbed increases that reduce the active water depth that seriously impacts the transportation of shallow draft water vessels through the waterway. Therefore, to avoid being stuck at any point of their navigational waterway, barges need to sacrifice their designed weight carrying capacity $w_b$. The maximum effective weight that a barge $b \in B$ can carry under this restriction would be $\overline{w}_{ijt\omega}$ where $\overline{w}_{ijt\omega} := \min\{w_{it\omega}, w_{ijt\omega}, w_{jt\omega}\}$ which is modelled as a stochastic parameter in our proposed formulation. With considering the water level fluctuation, in our model we additionally consider the possible delays experienced by towboats in the lock. Parameters $\Delta$, $l_{ij}$, $d_{ij}$, and $\overline{t}_{ij}$ are defined for this purpose that respectively represent the average delay in the number of locks between each origin-destination port, the traveling distance between each pair of ports $(i,j) \in (I,J)$, and the feasible travel time window. The average speed of towboat s is denoted as $\overline{v}_{st}$ and $t_l$ and $t_u$ represent the barge loading and unloading time. Finally, we consider that the commodity demand at destination port $j$, $d_{mjt}$, can be satisfied through waterway transportation or from any external supply chain with a penalty cost of $p_{mjt}$.

Our proposed model is a two-stage stochastic programming model where first-stage includes decision variables $\{Y_{sijt}\}_{\forall s \in S, i \in I, j \in J_i, t \in T}$ and $\{Y_{mbsijt}\}_{\forall m \in M, b \in B, s \in S, i \in I, j \in J_i, t \in T}$ that respectively determine which towboat to use between any origin-destination pair and which barge to use for carrying any particular product between any given origin and destination ports in a given time period. Second-stage decision variables $\{X_{mbsijt\omega}\}_{\forall m \in M, b \in B, s \in S, i \in I, j \in J_i, t \in T, \omega \in \Omega}$ denote the amount of commodities of type $m$ transported using barge $b$ of towboat $s$ along arc $(i,j)$ at time period $t$ under scenario $\omega$; $\{Z_{mit\omega}\}_{\forall m \in M, i \in I, t \in T, \omega \in \Omega}$ represent the amount of commodity m processed at port $i$ at time period $t$ under scenario $\omega$; $\{H_{mit\omega}\}_{\forall m \in M, i \in I \cup J, t \in T, \omega \in \Omega}$ denote the amount of commodities of type $m$ stored in port $i \in I \cup J$ at time period $t$ under scenario $\omega$; and $\{U_{mjt\omega}\}_{\forall m \in M, j \in J, t \in T, \omega \in \Omega}$ denote the amount of commodities of type $m$ shortage in destination port $j$ at time period $t$ under scenario $\omega$. It is worth mentioning that in inland waterway transportation, barges and towboats are provided by a third-party company, hence, in the first stage we determine the required number of barges and towboats to support the transportation, and the rest of decisions would be made in next stage. Our proposed model [IWT] can be seen as follows:

$$[\textbf{IWT}] \text{ Minimize} \sum_{s \in S} \sum_{i \in I} \sum_{j \in J_i} \sum_{t \in T} \left( \psi_{st} Y_{sijt} + \sum_{m \in M} \sum_{b \in B} \eta_{mbt} Y_{mbsijt} \right) + \quad (1)$$

$$\sum_{\omega \in \Omega} \sum_{t \in T} \sum_{m \in M} \rho_\omega \left( \sum_{i \in I \cup J} h_{mit} H_{mit\omega} + \sum_{b \in B} \sum_{s \in S} \sum_{(i,j) \in (I,J)} c_{mbsijt\omega} X_{mbsijt\omega} + \sum_{i \in I} \gamma_{mit} Z_{mit\omega} + \sum_{j \in J} \pi_{mjt} U_{mjt\omega} \right)$$

Subject to

$$\sum_{m \in M} Y_{mbsijt} \leq 1 \qquad \forall b \in B, s \in S, i \in I, j \in J_i, t \in T \quad (2)$$

$$\underline{\delta}_s Y_{sijt} \leq \sum_{m \in M} \sum_{b \in B} Y_{mbsijt} \leq \overline{\delta}_s Y_{sijt} \qquad \forall s \in S, i \in I, j \in J_i, t \in T \quad (3)$$

$$\sum_{m \in M} \sum_{b \in B} \sum_{s \in S} \sum_{j \in J_i} Y_{mbsijt} \leq \theta_{it} \qquad \forall i \in I, t \in T \quad (4)$$

$$\sum_{j \in J_i} \sum_{s \in S} Y_{sijt} \leq \tau_{it} \qquad \forall i \in I, t \in T \quad (5)$$

$$\sum_{m \in M} \sum_{s \in S} Y_{mbsijt} \leq a_{bit} \qquad \forall b \in B, i \in I, j \in J_i, t \in T \quad (6)$$

$$\sum_{j \in J_i} Y_{sijt} \leq a_{sit} \qquad \forall s \in S, i \in I, t \in T \quad (7)$$

$$\sum_{m \in M} \sum_{b \in B} (t_l + t_u) Y_{mbsijt} + \left( \frac{d_{ij}}{v_{st}} + \Delta l_{ij} \right) Y_{sijt} \leq \overline{t}_{ij} \qquad \forall s \in S, i \in I, j \in J_i, t \in T \quad (8)$$

$$Z_{mit\omega} \leq \phi_{mit\omega} \qquad \forall m \in M, i \in I, t \in T, \omega \in \Omega \quad (9)$$

$$Z_{mit\omega} + (1-\alpha_m)H_{mi,t-1,\omega} = \sum_{b\in B}\sum_{s\in S}\sum_{j\in J_i} X_{mbsijt\omega} + H_{mit\omega} \qquad \forall\, m \in M, j \in J, t \in T, \omega \in \Omega \qquad (10)$$

$$\sum_{b\in B}\sum_{s\in S}\sum_{i\in I} X_{mbsijt\omega} + (1-\alpha_m)H_{mj,t-1,\omega} = d_{mjt} + H_{mjt\omega} - U_{mjt\omega} \qquad \forall\, m \in M, j \in J, t \in T, \omega \in \Omega \qquad (11)$$

$$\sum_{m\in M} H_{mit\omega} \leq \overline{h}_i \qquad \forall\, i \in I \cup J, t \in T, \omega \in \Omega \qquad (12)$$

$$X_{mbsijt\omega} \leq \min\{\overline{w}_{ijt\omega}, \overline{w}_b\} Y_{mbsijt} \qquad \forall m \in M, b \in B, s \in S, i \in I, j \in J_i, t \in T, \omega \in \Omega \qquad (13)$$

$$X_{mbsijt\omega} \leq \rho_m v_b Y_{mbsijt} \qquad \forall m \in M, b \in B, s \in S, i \in I, j \in J_i, t \in T, \omega \in \Omega \qquad (14)$$

$$Y_{mbsijt}, Y_{sijt} \in \{0,1\} \qquad \forall m \in M, b \in B, s \in S, i \in I, j \in J_i, t \in T \qquad (15)$$

$$X_{mbsijt\omega}, H_{mit\omega}, H_{mjt\omega}, Z_{mit\omega} \in R^+ \qquad \forall m \in M, b \in B, s \in S, i \in I, j \in J_i, t \in T, \omega \in \Omega \qquad (16)$$

The objective function (1) sums up the first-stage costs and the expected second-stage costs. The first two terms in (1) represent the fixed towboat usage costs and fixed barge loading and unloading costs. Next four terms in objective (1) stands for commodity storage cost at the source and destination ports, commodity transportation cost through the network, commodity processing cost at any origin port, and the commodity shortage cost at any destination port. Constraints (2)-(7) are associated with the first stage. Constraints (2) state that at any given time period $t \in T$, one and only one commodity $m \in M$ can be loaded to a given barge $b \in B$. Constraints (3) are the towboat capacity constraints that restrict the minimum ($\underline{\delta}_s$) and maximum ($\overline{\delta}_s$) number of barges connected with a given towboat $s \in S$ at any time period $t \in T$ according to the capacity of that given towboat $s$. Constraints (4) and (5) respectively bound the barge and towboat usage at any given port according to the total barge availability ($\theta_{it}$) and towboat availability ($\tau_{it}$) of that port $i$ at that at that given period $t$. Further, due to aging and other related issues, barges and towboats needs to have periodic maintenance at different time periods of the year. If such activity occurs for any barge $b \in B$ or towboat $s \in S$ at any time $t \in T$ the respective barge and towboat become unavailable at that time period. This unavailability issue is captured through binary parameters $a_{sit}$ and $a_{bit}$ in constraints (6) and (7). Additionally, we impose the total travel time restriction for a towboat $s \in S$ between each origin destination port $(i,j) \in (I,J)$ at any specific time period $t \in T$ through constraints (8). Constraints (9) restrict commodity processing of any origin port $i \in I$ at time period $t \in T$ under scenario $\omega \in \Omega$ following stochastic commodity availability $\phi_{mit\omega}$ for commodity $m \in M$ in that respective port $i$. Constraints (10) and (11) are the flow balance constraints. The former ones ensure that at any given time $t \in T$, all processed commodity $m \in M$ will be either stored or transported in an origin or a destination port $i \in I \cup J$. Next constraints (11) state that at any given time period the incoming product or available inventory can be used to satisfy the commodity demand of the destination port and any surplus amount of commodity can be stored to the destination port inventory. Also, through (11) the model provide option of using any external supply source to satisfy the demand ($d_{mjt}$) or a portion of demand at any port $j \in J$ by paying a higher penalty cost of $p_{mjt}$. The inventory storage restriction at any port $i \in I \cup J$ is imposed trough constraints (12). Further, constraints (13) and (14) ensure that while using any barge $b \in B$ at any port $i \in I$ to transport commodity $m \in M$, the amount of commodity loaded to the particular barge should not exceed the weight (constraints (13)) and volumetric (constraints (14)) capacity of that barge $b$. Note that constraints (13) additionally capture the dredging impact showing that at each time period $t \in T$, a barge $b \in B$ is restricted to carry at most the minimum of $\{\overline{w}_{ijt\omega}, \overline{w}_b\}$ amount of commodity between each origin-destination port pair. Finally, constraints (15) are integrality constraints and constraints (16) represent the standard non-negativity constraints.

## 3. Computational Study and Managerial Insights
### 3.1 Solution Approach
Setting $|\Omega| = |T| = |S| = |B| = 1$, Model [IWT] can be reduced as a variant of fixed charge network flow problem which is already known to be an NP-hard problem [1]. Therefore, the full model [IWT] is also NP-hard and solving any large size instance this problem is very challenging for commercial solvers such as Gurobi, CPLEX. To overcome this challenge, we propose a parallelized hybrid decomposition algorithm based on Nested Decomposition Algorithm embedded with a sample average approximation algorithm [13] and a Progressive Hedging Algorithm[14], to solve the model to optimality (or near-optimality) in a reasonable timeframe. All algorithms are coded in python 2.7 and as an optimization solver we used Gurobi Optimizer 7.0.2. The algorithmic framework used for this problem can be seen in Figure 2(a). In Figure 2(a), N represent the number of available computer processors.

### 3.2 Experimental Results

In this study we investigate a real-life inland waterway transportation network including a total of thirteen inland waterway ports along the Mississippi River. Port of Rosedale, Greenville, Vicksburg, Natchez, Yazoo County, Geismar Louisiana, Greater Baton Rouge, South Louisiana, Gramercy, Memphis, Pemiscot County, New Madrid County, and the Port of Little Rock are selected for this purpose. These ports are directly connected to each other via the Mississippi River. Four commodities, rice, corn, woodchips, and fertilizer with an annual supply of 6.3 and 113.8, 8.3 and 0.4 million tons, respectively are selected to be transported from any of their origin ports to destination ports via the network under consideration. The annual demand of these commodities are 3.8, 68.3, 8.3, and 0.37 million tons. The barge and towboat transportation costs are obtained from [15]. Except these parameters, all other parameters are obtained from multiple research papers, online resources, and from ports by consultation with port personnel.

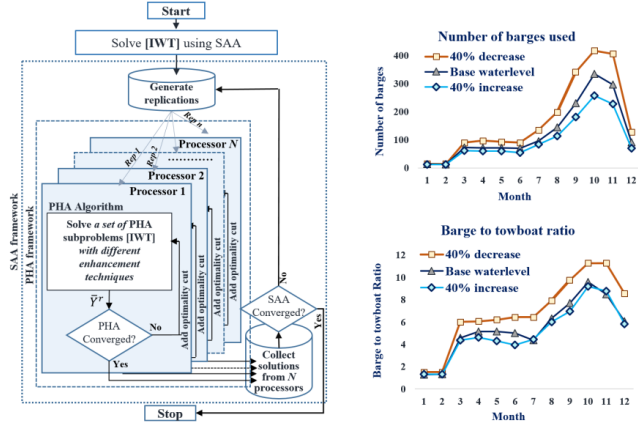

(a) Solution approach        (b) barge usage and barge to tow-boat ratio

Figure 2: Solution approach and case study results

As discussed earlier, the most serious problem faced by inland waterways are frequent water level fluctuations. Therefore, to investigate the impact of water level fluctuation on overall system performance we run sensitivity analysis with the realistic data generating two additional water level fluctuation scenarios with the base case. These scenarios are generated considering $\pm 40\%$ change in mean water level ($\overline{w}_{ijt\omega}$). Figure 2(b) summarize the experimental results where $t = 1$ stands for a representative day of month January, and the following 11 months are represented in an ascending order of months. The results show that with 40% increase in mean $\overline{w}_{ijt\omega}$, the number of barge usage for all time periods drops by 20% and this amount rise 34% when $\overline{w}_{ijt\omega}$ is changed by $-40\%$ from the base case. The peak barge usage is seen in month of October ($t = 10$) when the water level drop is most severe. Additionally, decrease in $\overline{w}_{ijt\omega}$ increase the towboat usage by 4%. On the other hand, 40% increase in mean water level w¯i jtw drops the towboat usage by 14%. Figure 2 also shows the result for barge to towboat ratio ($Y_{mbsijt}/Y_{sijt}$). This measure increases with the reduction in mean water level. Specifically, with 40% drop in $\overline{w}_{ijt\omega}$ this ratio reaches to a maximum of 12 barges per towboat (at time period 10, and 11). Opposite trend is observed for 40% increase case. But, with increased water level, barges now can carry more weights; therefore, we observe a reduction in barge to towboat ratio. Between January and July water level drop is not severe, therefore, decision variables do not show much variations in this time period. Additionally, we test the performance of our proposed algorithm. The results indicate that with incorporation of parallelization techniques on average about 53.9% of the solution time can be saved compared to that of the same algorithmic framework without parallelization procedure.

## 4. Conclusion and Future Research Directions

This paper proposes a two-stage stochastic MILP model to design and manage an inland waterway transportation-based logistics network while stochastic nature of commodity supply and water-level fluctuations are taken into consideration. The model jointly optimize towboat and barge assignment decisions and different supply chain decisions (e.g., inventory management, transportation decisions) under stochastic environment in such a way that the overall system cost can be minimized. A parallelized hybrid decomposition algorithm is proposed to solve the model. Results indicate that the presented parallelization schemes with hybrid decomposition algorithm can efficiently solve our proposed optimization model in a timely manner. A few Southeast US States were used as a testbed to visualize and validate the modeling results. Results reveal the impact of uncertain water level fluctuation on the inland waterway transportation network. The managerial insights obtained from this study may help decision makers to design and manage a cost-efficient inland waterway-based supply chain network under uncertainty. This study can be extended by incorporating detailed consideration of barge and tow routing, scheduling, and re-positioning issues impact the inland waterway port operations. Next, the impacts of port operations under both natural (e.g., hurricane, tornado) and/or human-induced (e.g., cyber attack)[16] disruptions can also be investigated. Future study will address all these issues.